\documentclass[aps,secnumarabic,pra,twocolumn,amsmath,amssymb,nofootinbib,eqsecnum,superscriptaddress,nobibnotes,floatfix]{revtex4-1}
\usepackage{float}
\usepackage[squaren]{SIunits}
\usepackage{verbatim}
\usepackage{mathrsfs}
\usepackage{stmaryrd}
\usepackage{comment}
\usepackage{braket}
\usepackage{bbold}
\usepackage[colorlinks]{hyperref}
\usepackage{xcolor}
\usepackage[latin1]{inputenc}
\usepackage{graphicx}
\usepackage{amsmath}
\usepackage{dsfont}
\usepackage{amsfonts,amssymb}
\usepackage{amsthm}
\usepackage{xspace}
\usepackage{times}
\usepackage{longtable}
\usepackage{extarrows}
\usepackage{ulem}   
\begin{document}

\title{Quantum simulation of Abelian Wu-Yang monopoles in spin-1/2 systems}
\date{\today}

\author{Ze-Lin Zhang}
\affiliation{Department of Physics, Fuzhou University, Fuzhou 350002, P. R. China}
\affiliation{Fujian Key Laboratory of Quantum Information and Quantum Optics, Fuzhou University, Fuzhou 350002, P. R. China}
\author{Ming-Feng Chen}
\affiliation{Department of Physics, Fuzhou University, Fuzhou 350002, P. R. China}
\affiliation{Fujian Key Laboratory of Quantum Information and Quantum Optics, Fuzhou University, Fuzhou 350002, P. R. China}
\author{Huai-Zhi Wu}
\affiliation{Department of Physics, Fuzhou University, Fuzhou 350002, P. R. China}
\affiliation{Fujian Key Laboratory of Quantum Information and Quantum Optics, Fuzhou University, Fuzhou 350002, P. R. China}
\author{Zhen-Biao Yang}
\email{E-mail address: zbyang@fzu.edu.cn. The author to whom any correspondence should be addressed.}
\affiliation{Department of Physics, Fuzhou University, Fuzhou 350002, P. R. China}
\affiliation{Fujian Key Laboratory of Quantum Information and Quantum Optics, Fuzhou University, Fuzhou 350002, P. R. China}

\begin{abstract}
  With the help of the Berry curvature and the first Chern number $($$\textit{C}_1$$)$, we both analytically and numerically investigate and thus simulate artificial magnetic monopoles formed in parameter space of the Hamiltonian of a driven superconducting qubit. The topological structure of a spin-1/2 system $($qubit$)$ can be captured by the distribution of Berry curvature, which describes the geometry of eigenstates of the Hamiltonian. Degenerate points in parameter space act as sources $($$\textit{C}_1$ = $1$, represented by quantum ground state manifold$)$ or sinks $($$\textit{C}_1$ = $-1$, represented by quantum excited state manifold$)$ of the magnetic field. We note that the strength of the magnetic field $($described by Berry curvature$)$ has an apparent impact on the quantum states during the process of topological transition. It exhibits an unusual property that the transition of the quantum states is asymmetric when the degenerate point passes from outside to inside and again outside the manifold spanned by system parameters. Our results also pave the way to explore intriguing properties of Abelian Wu-Yang monopoles in other spin-1/2 systems.
\end{abstract}

\pacs{}

\maketitle
\section{Introduction}
\label{sec: Introduction}
  In nature, magnetic poles always come in twos, a north and a south. Yet their electrostatic cousins, positive and negative charges, exist independently. In 1931, Dirac developed a theory of monopoles consistent with both quantum mechanics and the gauge invariance of the electromagnetic field~\cite{PAMD-1931}. The existence of a single Dirac monopole would not only address this seeming imbalance which appears in the Maxwell's equations, but would also explain the quantization of electric charge~\cite{PAMD-1931,PAMD-1948}. Up to now, magnetic monopole analogues have been created in many different ways, such as superfluid $^3$He~\cite{SB-1976,MS-1987}, exotic spin ice~\cite{CMS-2008,BGCAPF-2009,LRPCB-2010} and spinor Bose-Einstein condensates ~\cite{Machida-1998,H-1998,PM-2009,RPM-2011,RRKMH-2014,RRTMH-2015,TRMHM-2016,MP-2013}. Methods in Refs.~\cite{RRKMH-2014,RRTMH-2015,TRMHM-2016} could be regarded as excellent examples of quantum simulation of magnetic monopoles. Quantum simulation was originally conceived by Feynman in 1982~\cite{RPF-1982}, which permits the study of quantum systems that are difficult to study in laboratory. For this reason, simulators are especially aimed at providing insight about the behavior of more inaccessible systems appearing in nature. By introducing the point-like topological defects accompanied with a vortex filament into the spin texture of a dilute Bose-Einstein condensate, researchers provided an ideal analogue to Dirac monopole~\cite{PM-2009}.

  The topological properties of quantum systems play an extraordinary role in our understanding of the fundamental significance of natural phenomena. For example, the first Chern number $($$\textit{C}_1$$)$~\cite{SSC-1946}, which is a kind of robust topological invariants staying the same by small perturbations to the system can be used to help categorize physical phenomena. It is closely related to Berry phase that arises in cyclic adiabatic evolution of a system  in addition to the dynamical counterpart~\cite{MVB-1984}. The point-like topological defects as with degeneracy points in Hamiltonian parameter space of a spin-1/2 system could be viewed as the physical counterpart of topological invariant, which can be described by the first Chern number~\cite{DCAJ-2004}. $\textit{C}_1$ can be extracted by integrating Berry curvature over the closed surface. Gritsev \textit{et al.}~\cite{VGAP-2012} proposed an effective method to measure the Berry curvature directly via the nonadiabatic response on physical observables to the rate of change of an external parameter. The method provides a powerful and generalized approach to explore topological properties in arbitrary quantum systems where the Hamiltonian can be written in terms of a set of externally controlled parameters. Taken into account this method, some researchers measured the topological transition $\textit{C}_1=1\rightarrow0$ in a single superconducting qubit~\cite{SKKSGVPPL-2014}, and others observed the topological transitions in interacting quantum circuits~\cite{PR-2014}. Experimental schemes have also been proposed to simulate the dynamical quantum Hall effect in a Heisenberg spin chain with interacting superconducting qubits~\cite{YZXYZ-2015}, and to realize several-spin one-dimensional Heisenberg chains using nuclear magnetic resonance $($NMR$)$ simulators~\cite{LLLNLPD-2016}.

  In this paper, we study the Wu-Yang monopoles \cite{YCN-1975}, which remove out the ``Dirac string'' by gauge transformation in parameter space of the Hamiltonian of a driven superconducting qubit for both geometry $($Berry curvature$)$ and topology $($the first Chern number, $\textit{C}_1$$)$. The topological structure of the qubit can be captured by the distribution of Berry curvature, which describes the geometry of eigenstates of the Hamiltonian. We note that degenerate points in parameter space of the Hamiltonian act as the sources $($sinks$)$ of $\textit{C}_1$ and are analogues to magnetic monopoles $\textit{g}_{\textit{N}(\textit{S})}$ $($$C_1$ = $1$ $\leftrightarrow$ $\textit{g}_{\textit{N}}$, $C_1$ = $-1$ $\leftrightarrow$ $\textit{g}_{\textit{S}}$$)$. We also note that the transition of quantum states is asymmetric during the process when the degeneracy passes from outside to inside and again outside the manifold spanned by system parameters, and the Berry curvature and the fidelity of quantum states have some interesting correlations during the process of topological transition. We give a preliminary explanation to it by introducing the notion of magnetic charges. This general method also can be simulated by other spin-1/2 systems. For example, it can be extended to that in an NMR system and is possible to experimentally investigate more intriguing properties of multi-monopoles, which could be used to construct new kinds of devices based on synthetic magnetic fields.

  The configuration of this paper proceeds as follows. In Sec.~\ref{sec: Geometry and topology in the specific state manifold}, we introduce the quantum geometric metric tensor and show its relation to the Berry curvature. In Sec.~\ref{sec: The Chern-Gauss-Bonnet theorem}, we describe how the first Chern numbers are obtained from Berry curvatures. In Sec.~\ref{sec: Measuring the Berry curvature}, we outline an effective method to measure the Berry curvature directly via the nonadiabatic response on physical observables to the rate of change of an external parameter. In Sec.~\ref{sec: The Chern-Gauss-Bonnet theorem}, we describe how the first Chern numbers are obtained from Berry curvatures. As a useful example, we introduce a physical model for the simulation of the Abelian Wu-Yang monopoles by a driven superconducting qubit in Sec.~\ref{sec: From Dirac monopole to Wu-Yang monopole}.  In Sec.~\ref{sec: Physical Model for Implementation} we explain how Wu-Yang monopoles are differ from the Dirac monopoles through two kinds of quantum state manifolds. Finally, in Sec.~\ref{sec: Results}, we discuss some interesting correlations between the Berry curvature and the quantum states during the process of topological transition, we then describe the experimental feasibility of this theoretical method.

\section{Geometry and topology in the specific state manifold}
\label{sec: Geometry and topology in the specific state manifold}

  Consider a family of parameter-dependent Hamiltonian ${\vec{\lambda}}$ for a quantum system and require $\vec{\lambda}$ to depend smoothly on a set of parameters $\vec{\lambda} = (\lambda^{\mathbb{1}}, \lambda^{\mathbb{2}}, \cdots)\in\mathcal{M}$ $(\mathcal{M}$ denotes the Hamiltonian parameters base manifold$)$ and act over the Hilbert space. The outline font $\mathbb{1}$ and $\mathbb{2}$ indicate different indices. The distance between the two neighbouring specific $($say, ground$)$ state wave functions $|\psi_{0}(\vec{\lambda})\rangle$ and $|\psi_{0}(\vec{\lambda} + \textit{d}\vec{\lambda})\rangle$ over $\mathcal{M}$ is~\cite{JPGV-1980,MCFL-2010,MKVGAP-2013}
  \begin{eqnarray}\label{distance}
  \textit{ds}^2 = 1 - |\langle \psi_{0}(\vec{\lambda})|\psi_{0}(\vec{\lambda} + \textit{d}\vec{\lambda})\rangle|^2 = \underset{\mu\nu}{\sum}\textit{g}_{\mu\nu}\textit{d}\lambda^{\mu}\textit{d}\lambda^{\nu},~~~
  \end{eqnarray}
  where the quantum $($Fubini-Study$)$ metric tensor $\textit{g}_{\mu\nu}$ associated with the ground state manifold is the symmetric real part of the quantum geometric tensor $\textit{Q}_{\mu\nu}$:
  \begin{equation}\label{quantum geometric tensor}
  \textit{Q}_{\mu\nu} = \langle\partial_{\mu}\psi_{0}|\partial_{\nu}\psi_{0}\rangle-\langle\partial_{\mu}\psi_{0}|\psi_{0}\rangle\langle \psi_{0}|\partial_{\nu}\psi_{0}\rangle,
  \end{equation}
  \begin{equation}\label{Fubini-Study metric tensor}
  \textit{g}_{\mu\nu} = \mathrm{Re}[\textit{Q}_{\mu\nu}] = (\textit{Q}_{\mu\nu}+\textit{Q}^\ast_{\mu\nu})/{2},
  \end{equation}
  with $\partial_{\mu(\nu)}\equiv{\partial}/{\partial\lambda^{\mu(\nu)}}$. The Hermitian metric tensor $\textit{Q}_{\mu\nu}$ remains unchanged under arbitrary $\lambda$-dependent $\textit{U}(1)$ local gauge transformation of $|\psi_{0}(\vec{\lambda})\rangle$. In another pioneering work~\cite{MVB-1984}, Berry introduced the concept of the geometric phase and the related geometric curvature $($also called Berry phase and Berry curvature$)$. The Abelian Berry curvature $\textit{F}_{\mu\nu}$ is given by the antisymmetric imaginary part of $\textit{Q}_{\mu\nu}$:
  \begin{eqnarray}\label{Berry curvature}
  \textit{F}_{\mu\nu} = -2\mathrm{Im}[\textit{Q}_{\mu\nu}] = \textit{i}(\textit{Q}_{\mu\nu}-\textit{Q}^\ast_{\mu\nu}) = \partial_{\mu}\textit{A}_{\nu}-\partial_{\nu}\textit{A}_{\mu},~~~~~
  \end{eqnarray}
  where $\textit{A}_{\mu(\nu)}=\textit{i}\langle \psi_{0}(\vec{\lambda})|\partial_{\mu(\nu)}|\psi_{0}(\vec{\lambda})\rangle$ is just the Berry connection.

  \subsection{The Chern-Gauss-Bonnet theorem}
  \label{sec: The Chern-Gauss-Bonnet theorem}

\begin{figure}[t]
    \centering
    \includegraphics[width=3in]{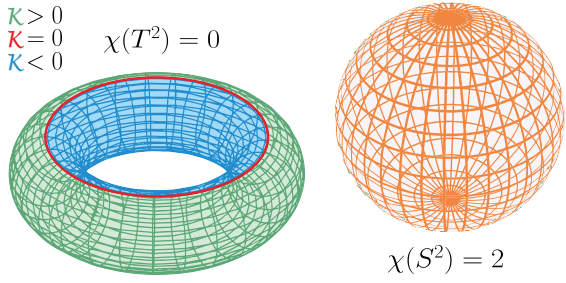}
    \caption{$($Color online$)$ Euler characteristic for a torus $($doughnut$)$ and a sphere. From the torus's point of view, the Gauss curvature is positive when the curving of the surface is elliptic $($the green area$)$. If the parabolic likes a plane $($the red circle$)$, then the Gauss curvature is zero. If the surface stars to show hyperbolic curving such as a saddle $($the blue area$)$, then the Gauss curvature becomes negative. From the sphere's point of view, the Gauss curvature is a positive constant. Intuitively, $\chi(T^2) = 2(1-\mathfrak{g}(T^2))=0$ and $\chi(S^2) = 2(1-\mathfrak{g}(S^2))=2$.}
    \label{fig:figure1}
  \end{figure}

  Let $\mathcal{M}^{m}$ be a compact oriented Riemann manifold of even dimension $($$\textit{m} = 2\textit{n}$$)$ and define on $\mathcal{M}^{m}$ a global \textit{m} form, the Chern-Gauss-Bonnet $($\textit{C-G-B}$)$ formula says that
  \begin{equation}\label{Chern-Gauss-Bonnet}
  \int_{\mathcal{M}^{m}} \textit{e}(\Omega) = \chi(\mathcal{M}),
  \end{equation}
  where $\textit{e}(\Omega)$ is the Euler class, $\chi(\mathcal{M})\equiv 2(1-\mathfrak{g})$ is the integer Euler characteristic describing the topology of the smooth manifold $\mathcal{M}$ and $\mathfrak{g}$ is the genus that also can be considered as the number of holes of the manifold. As shown in Fig.~\ref{fig:figure1}, two simplest closed manifolds are taken for example. In the lower dimensional version, the $\textit{C-G-B}$ theorem reduces to the Gauss-Bonnet $($\textit{G-B}$)$ theorem. The Fubini-Study tensor $\textit{g}_{\mu\nu}$ defines a Riemannian manifold related to the ground state. Especially , the structure of the Riemannian manifold provides a different topological integer, given by using the $\textit{G-B}$ theorem to the metric tensor in quantum version~\cite{PP-2006}:
  \begin{eqnarray}\label{Gauss-Bonnet}
  \frac{1}{2\pi}\bigg(\iint_{\mathcal{M}}\textit{K}~\textit{dS} + \oint_{\partial{\mathcal{M}}}\kappa_{\textit{g}}~\textit{dl}\bigg) = \chi(\mathcal{M}),
  \end{eqnarray}
  where $\textit{K}$ $($Gauss curvature$)$, $\textit{dS}$ $($area element$)$, $\kappa_{\textit{g}}$ $($geodesic curvature$)$, and $\textit{dl}$ $($line element$)$ are geometric invariants, meaning that they remain unchanged under any change of variables. The left side of Eq.~$($\ref{Gauss-Bonnet}$)$ are the bulk $($$\mathcal{M}$$)$ and boundary $($$\partial{\mathcal{M}}$$)$ contributions to $\chi(\mathcal{M})$ of the Riemannian manifold. If the manifold $\mathcal{M}$ is compact and without boundary $($closed$)$, then the boundary Euler integrals vanish, as we prove in detail in Appendix~{I} of the Supplementary data. In this paper we will focus only on the two-dimensional $($$\textit{m}$ = $2$ in Eq.~$($\ref{Chern-Gauss-Bonnet}$)$$)$ version and the dimensionality here is that of parameter space $($i.e., ${\mathcal{S}}^2$$)$ which is composed by the polar angle $\theta$ and the azimuthal angle $\phi$ of a magnetic field applied to a spin-1/2 system. Then we get the global $\textit{G-B}$ theorem on the sphere
  \begin{eqnarray}\label{Global Gauss-Bonnet}
  \frac{1}{2\pi}\oint_{{\mathcal{S}}^2}\textit{K}~\textit{dS} = \chi({\mathcal{S}}^2).
  \end{eqnarray}

  To catch the significance of the first Chern number $\textit{C}_1$, we need to adiabatically change these parameters around a loop that bounds a sphere ${\mathcal{S}}^2$ to acquire a Berry phase, which can be written as
  \begin{eqnarray}\label{quantum geometric tensor relation formula}
  \varphi_{\textit{Berry}} = \iint_{{\mathcal{S}}^2}\textit{F}_{\mu\nu}\textit{dS}_{\mu\nu} = \iint_{{\mathcal{S}}^2}\vec{\textit{F}}\cdot\textit{d}\vec{\textit{S}},
  \end{eqnarray}
  where $\textit{dS}_{\mu\nu}$ is a directed surface element, $\vec{\textit{S}}$ is a vector normal to the sphere ${\mathcal{S}}^2$ and $\vec{\textit{F}}$ is a vector known as the Berry curvature analogous to the magnetic field in electromagnetism, which is given by the off-diagonal components of the electromagnetic tensor $\textit{F}_{\mu\nu}$, see in Eq.~$($\ref{Berry curvature}$)$. For example, the Berry curvatures $\textit{F}^{(N)}_{\theta\phi}$ and $\textit{F}^{(S)}_{\theta\phi}$ only have off-diagonal components.

  As we all know by now, Berry phase depends on the  $\textit{U}(1)$ local gauge choice $|\psi_{i}\rangle\rightarrow \textit{e}^{i\varphi(\theta,\phi)}|\psi_{i}\rangle$, where $|\psi_{i}\rangle$ is a certain eigenstate in this paper $($subscript \textit{i} = 0, 1$)$, showing that the Berry curvature is gauge invariant. Therefore, we obtain the integral
  \begin{eqnarray}\label{the first Chern number}
  \textit{C}_1 = \frac{1}{2\pi}\oint_{{\mathcal{S}}^2}\textit{F}_{\mu\nu}\textit{dS}_{\mu\nu} = \frac{1}{2\pi}\oint_{{\mathcal{S}}^2}\vec{\textit{F}}\cdot\textit{d}\vec{\textit{S}},
  \end{eqnarray}
  is a kind of robust topological invariant known as the first Chern number, and it could be viewed as counting the number of times an eigenstate circles around a sphere in the Hilbert space~\cite{SKKSGVPPL-2014}.

  \subsection{Measuring the Berry curvature}
  \label{sec: Measuring the Berry curvature}
  In analogy to electrodynamics, the local gauge-dependent Berry connection $\textit{A}_{\mu}$ can never be physically observed, while Berry curvature $\textit{F}_{\mu\nu}$ is gauge-invariant and may be related to a physical observable that manifests the local geometric property of the eigenstates in the parameter space. The first Chern number reveals the global topological property of such a Hamiltonian manifold. In fact, $\textit{C}_1$ exactly counts the number of degenerate points enclosed by parameter space $\mathcal{S}^2$, see in Appendix~{II} of the Supplementary data, where we endow it with physical meaning by using the conception of the magnetic monopole. We substitute $\textit{A}_{\mu}$ into $\textit{F}_{\mu\nu}(\vec{\textit{F}})$ and rewrite the Berry curvature as
  \begin{eqnarray}\label{large Berry curvature}
  \textit{F}_{\mu\nu} = i\sum_{n\neq0}\frac{\langle \psi_{0}|\partial_{\mu}\hat{H}|\psi_{n}\rangle\langle \psi_{n}|\partial_{\nu}\hat{H}|\psi_{0}\rangle -(\nu\leftrightarrow\mu)}{(\textit{E}_n-\textit{E}_0)^2},~~~~~~~~~
  \end{eqnarray}
  where $\textit{E}_n$ and $|\psi_{n}\rangle$ are the $\textit{n}$-th eigenvalue and its corresponding eigenstate of the Hamiltonian $\hat{H}$, respectively. Eq.~$($\ref{large Berry curvature}$)$ indicates that degeneracies are some singularities that will contribute nonzero terms to $\textit{C}_1$ in Eq.~$($\ref{the first Chern number}$)$.

  In order to extract the Chern number of closed manifolds in the parameter space of the two-level system Hamiltonian, we analytically describe a simple topological structure of a superconducting qubit driven by a microwave field. In Ref.~\cite{VGAP-2012}, it states that Berry curvature can be extracted from the linear response of the qubit to nonadiabatic manipulations of its Hamiltonian $\hat{H}$$($$\mu = \theta$, $\nu = \phi$$)$, which leads to a general force $\textit{M}_\phi\equiv-\langle\psi_{0}(t)|\partial_{\phi}\hat{H}|\psi_{0}(t)\rangle$, given by ~\cite{VGAP-2012,MVBJMR-1993,SKKSGVPPL-2014}
  \begin{equation}\label{general_force}
  \textit{M}_\phi = \textrm{const} + \upsilon_\theta\textit{F}_{\theta\phi}+\mathcal{O}(\upsilon^2),
  \end{equation}
  where $\upsilon_\theta$ is the rate of change of the parameter $\theta$ $($quench velocity$)$ and $\textit{F}_{\theta\phi}$ is a component of the Berry curvature tensor. To neglect the nonlinear term, the system parameters should be ramped slowly enough or quasi-adiabaticly.

\section{From Dirac monopole to Wu-Yang monopole}
\label{sec: From Dirac monopole to Wu-Yang monopole}

\begin{figure}[t]
    \centering
    \includegraphics[width=3 in]{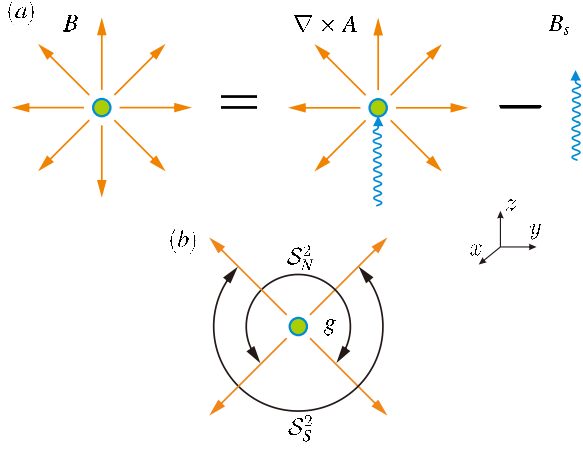}
    \caption{$($Color online$)$ From Dirac monopole to Wu-Yang monopole. (a) Dirac monopole. Maxwell's equations can accommodate magnetic monopoles, due to quantum mechanics, it is always possible to create a magnetic field emerging from a point by importing the field from far distance to the point through an infinitely thin physically undetectable magnetic flux tube, which is called the Dirac string. From the endpoint of the string, magnetic field lines emerge radially outwards in the same way as electric field lines emerge from an electric point charge, so that the endpoint acts as a magnetic monopole. (b) Wu-Yang monopole. By selecting different coordinate systems to eliminate the singularity of Dirac string.}\label{Dirac_Wu_monopole}
    \label{fig:figure2a_2b}
  \end{figure}

   In order to discuss in more detail about Dirac monopole, we first consider a monopole with the magnetic field sitting at the origin
  \begin{eqnarray}\label{div_B}
  \nabla\cdot \vec{\textit{B}}= 4\pi\textit{g}\delta(\vec{\textit{r}}).
  \end{eqnarray}
  It follows from $\nabla^2(1/\textit{r})=-4\pi\delta(\vec{\textit{r}})$ and $\nabla(1/\textit{r})=-\vec{\textit{r}}/{{\textit{r}}^3}$ that the solution of this equation is
  \begin{eqnarray}\label{B}
  \vec{\textit{B}}= \vec{\textit{F}}(r,\theta,\phi) = {\textit{g}\vec{\textit{r}}}/{{\textit{r}}^3},
  \end{eqnarray}
  where $\textit{g} = \mp 1/2$. The magnetic flux $\Phi$ is obtained by integrating over a sphere $\mathcal{S}^2$ of radius \textit{r} so that
  \begin{eqnarray}\label{Phi}
  \Phi = \oint_{\mathcal{S}^2}\vec{\textit{B}}\cdot\textit{d}\vec{\textit{S}} = 4\pi\textit{g}.
  \end{eqnarray}
  But if $\vec{\textit{B}} = \nabla\times\vec{\textit{A}}$, this integral would have to vanish. Thus magnetic vector potential $\vec{\textit{A}}$ cannot exist everywhere on $\mathcal{S}^2$, even though $\nabla\cdot \vec{\textit{B}}$ is only non-zero at the origin, and the best we can do is to find an $\vec{\textit{A}}$ defined everywhere except on a line joining the origin to infinity, such that $\vec{\textit{B}} = \nabla\times\vec{\textit{A}}$. To see this is possible, it may reasonably consider the field due to an infinitely long and thin solenoid placed along the negative \textit{z} axis with its positive pole which has strength \textit{g} at the origin~\cite{PGDIO-1978}. For example, let us introduce the singular vector potential
  \begin{eqnarray}\label{singular vector potential}
  \textit{A}_r = \textit{A}_{\theta} = 0,~~~~\textit{A}_{\phi} = \frac{\textit{g}(1-\cos\theta)}{\textit{r}\sin\theta},
  \end{eqnarray}
  and verify that
  \begin{eqnarray}\label{curl_A}
  \nabla\times \vec{\textit{A}} = {\textit{g}\vec{\textit{r}}}/{{\textit{r}}^3} + \vec{\textit{B}}_s,
  \end{eqnarray}
  where $\vec{\textit{B}}_s$ is the singular vector field along \textit{z}-axis, with the expression
  \begin{eqnarray}\label{singular vector field}
  \vec{\textit{B}}_s=
  \begin{cases}
  4\pi\textit{g}\delta(\textit{x})\delta(\textit{y})\theta(\textit{z}),&\textit{z}<0,\theta=\pi \\
  0,&\textit{z}>0,\theta = 0 \\
  \end{cases}.
  \end{eqnarray}
  The singularity along the \textit{z}-axis is called the Dirac string and reflects the poor choice of the coordinate system, as is shown in Fig.~\ref{fig:figure2a_2b}$($a$)$. This magnetic field differs from $\vec{\textit{B}}$ only by the singular magnetic flux along the solenoid but it is clearly source-free; while at the origin, $\vec{\textit{B}}$ vanishes. Thus it may be
  represented by a vector potential, $\vec{\textit{A}}$ $($say$)$, everywhere and we may write
  \begin{eqnarray}\label{curl_B2}
  \vec{\textit{B}} = \nabla\times \vec{\textit{A}} - \vec{\textit{B}}_s.
  \end{eqnarray}

  Now, let us describe how these monopoles differ from the standard Dirac monopoles. Under the condition of quantum excited state manifold, and from Eq.~$($3$)$ and Eq.~$($4$)$ in Appendix~{I} of the Supplementary data, we obtain the magnetic field of the south monopole
  \begin{eqnarray}\label{Berry curvature2}
  \textit{F}_{\theta\phi}^{(S)}
  = -2\mathrm{Im}[\textit{Q}_{\theta\phi}]
  =\frac{1}{2}
  \left( \begin{array}{cc}
  0 & -\sin\theta\\
  \sin\theta & 0\\
  \end{array}
  \right).
  \end{eqnarray}
  The corresponding Berry curvature $\vec{\textit{F}}^{(S)} = -1/2\sin\theta \textit{d}\theta\wedge\textit{d}\phi$ is a symplectic form on ${\mathcal{S}}^2$. If we transform it to the Coulomb-like magnetic field
  \begin{eqnarray}\label{field strength}
  \vec{\textit{F}}^{(S)}(r,\theta,\phi) = {\textit{g}_{\textit{S}}\vec{\textit{r}}}/{{\textit{r}}^3} = -{\vec{\textit{r}}}/{2{\textit{r}}^3},
  \end{eqnarray}
  it turns out to be the magnetic field originating from a monopole located at the origin with magnetic charge $\textit{g}_{\textit{S}} = -1/2$~\cite{DCAJ-2004}.
  Similarly, if we take another eigenstate which corresponds to the quantum ground state manifold
  \begin{eqnarray}\label{qubit2}
  |\psi_{0}(\theta, \phi)\rangle = -\sin({\theta}/{2})|0\rangle + \textit{e}^{i\phi}\cos({\theta}/{2})|1\rangle,
  \end{eqnarray}
  where we set $\sin({\theta}/{2}) = -\frac{\Omega}{2}\big/{\sqrt{\frac{\Omega^2}{4} + (\textit{E}_0 - \frac{\Delta}{2})^2}}$, and $\cos({\theta}/{2}) = -(\textit{E}_0 - \frac{\Delta}{2})\big/{\sqrt{\frac{\Omega^2}{4} + (\textit{E}_0 - \frac{\Delta}{2})^2}}$, then we have the magnetic field of the north monopole
  \begin{eqnarray}\label{Berry curvature3}
  \textit{F}^{(N)}_{\theta\phi}&=&\frac{1}{2}
  \left( \begin{array}{cc}
  0 & \sin\theta\\
  -\sin\theta & 0\\
  \end{array}
  \right).~
  \end{eqnarray}
  The corresponding Berry curvature is $\vec{\textit{F}}^{(N)} = 1/2\sin\theta \textit{d}\theta\wedge\textit{d}\phi$, and the magnetic field
  \begin{eqnarray}\label{field strength2}
  \vec{\textit{F}}^{(N)}(r,\theta,\phi) = \textit{g}_{\textit{N}}\vec{\textit{r}}/{{\textit{r}}^3} = {\vec{\textit{r}}}/{2{\textit{r}}^3},
  \end{eqnarray}
  with the magnetic charge $\textit{g}_{\textit{N}} = 1/2$.

  T. T. Wu and C. N. Yang~\cite{YCN-1975} noticed that it may employ more than one vector potential to describe monopoles. For example, we may avoid singularities if we adopt $\vec{\textit{A}}_{\textit{N}}$ in the northern hemisphere and $\vec{\textit{A}}_{\textit{S}}$ in the southern hemisphere of the sphere ${\mathcal{S}^2}$ surrounding the monopole,
  as depicted in Fig.~\ref{fig:figure2a_2b}$($b$)$. It shows that the vector potential $\vec{\textit{A}}_{\textit{N}}$ in region of ${\mathcal{S}_{\textit{N}}^2}$ can be expressed as
  \begin{eqnarray}\label{singular vector potential}
  \begin{split}
  (\textit{A}_r)_{\textit{N}} = (\textit{A}_{\theta})_{\textit{N}} = 0,~~~~~
  (\textit{A}_{\phi})_{\textit{N}} = \frac{\textit{g}(1-\cos\theta)}{\textit{r}\sin\theta},
  \end{split}
  \end{eqnarray}
  and the vector potential $\vec{\textit{A}}_{\textit{S}}$ in region of ${\mathcal{S}_{\textit{S}}^2}$ can be expressed as
  \begin{eqnarray}\label{singular vector potential2}
  \begin{split}
  (\textit{A}_r)_{\textit{S}} = (\textit{A}_{\theta})_{\textit{S}} = 0,~~~~~
  (\textit{A}_{\phi})_{\textit{S}} = -\frac{\textit{g}(1+\cos\theta)}{\textit{r}\sin\theta}.
  \end{split}
  \end{eqnarray}
  Obviously, the two vector potentials yield the magnetic field $\vec{\textit{B}}$ = $\textit{g}\vec{\textit{r}}/{{\textit{r}}^3}$, which is non-singular everywhere on the sphere~\cite{MN-1998}.

  Need of special note is that the magnetic monopoles we simulate here are the Abelian Wu-Yang monopoles, which by selecting different coordinate systems to eliminate the singularity of Dirac string. The two coordinate systems are characterized by the choice of two different Berry curvatures, see more in Appendix~{II} of the Supplementary data.

\section{Physical Model for Implementation}
\label{sec: Physical Model for Implementation}

  \begin{figure}[h]
    \centering
    \includegraphics[width=3in]{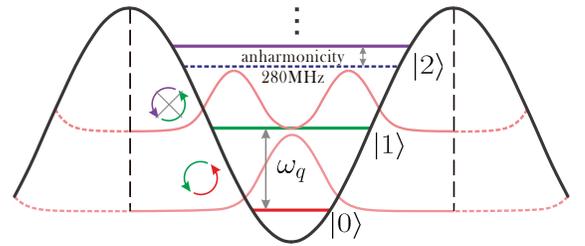}
    \caption{$($Color online$)$ Energy spectrum of the superconducting transmon qubit. Here we assume the qubit is effectively a nonlinear resonator, with a transition frequency of $\omega_q = 4.395$ GHz and the anharmonicity of 280 MHz, to ensure that the qubit transition only occurs between the ground state and the first excited state~\cite{KYGHSMBDGS-2007}.}
    \label{fig:figure3}
  \end{figure}

  As we have mentioned above, the degenerate points emerging from the Berry curvature $\textit{F}_{\mu\nu}$ act as the sources $($the north magnetic charge $\textit{g}_{\textit{N}}$$)$ and sinks $($the south magnetic charge $\textit{g}_{\textit{S}}$$)$ of $\textit{C}_1 (\pm 1)$ and are analogous to Wu-Yang monopoles in parameter space. We reconsider the proposal that use a superconducting transmon qubit described in Ref.~\cite{SKKSGVPPL-2014}. As seen in Fig.~\ref{fig:figure3}, where an anharmonicity of 280 MHz makes the qubit an effective two-level system in the parameter scope. In the rotating frame of a microwave drive with frequency $\omega_m$, the Hamiltonian for the qubit can be written as $($$\hbar\equiv$1$)$~\cite{FGCBV-2002,KYGHSMBDGS-2007}
  \begin{eqnarray}\label{Hamiltonian}
  \hat{H}
  &=& 1/{2}[\Delta\hat{\sigma}_z+\Omega\hat{\sigma}_x\cos\phi+\Omega\hat{\sigma}_y\sin\phi],
  \end{eqnarray}
  where $\Delta = \omega_m-\omega_q$, $\hat{\sigma}_i (i =x,y,z)$ is the Pauli spin matrix, $\phi$ and $\Omega$ are the phase of the drive tone and the amplitude of the drive tone as the Rabi frequency, respectively.
  \begin{figure}[t]
    \centering
    \includegraphics[width=3.2in]{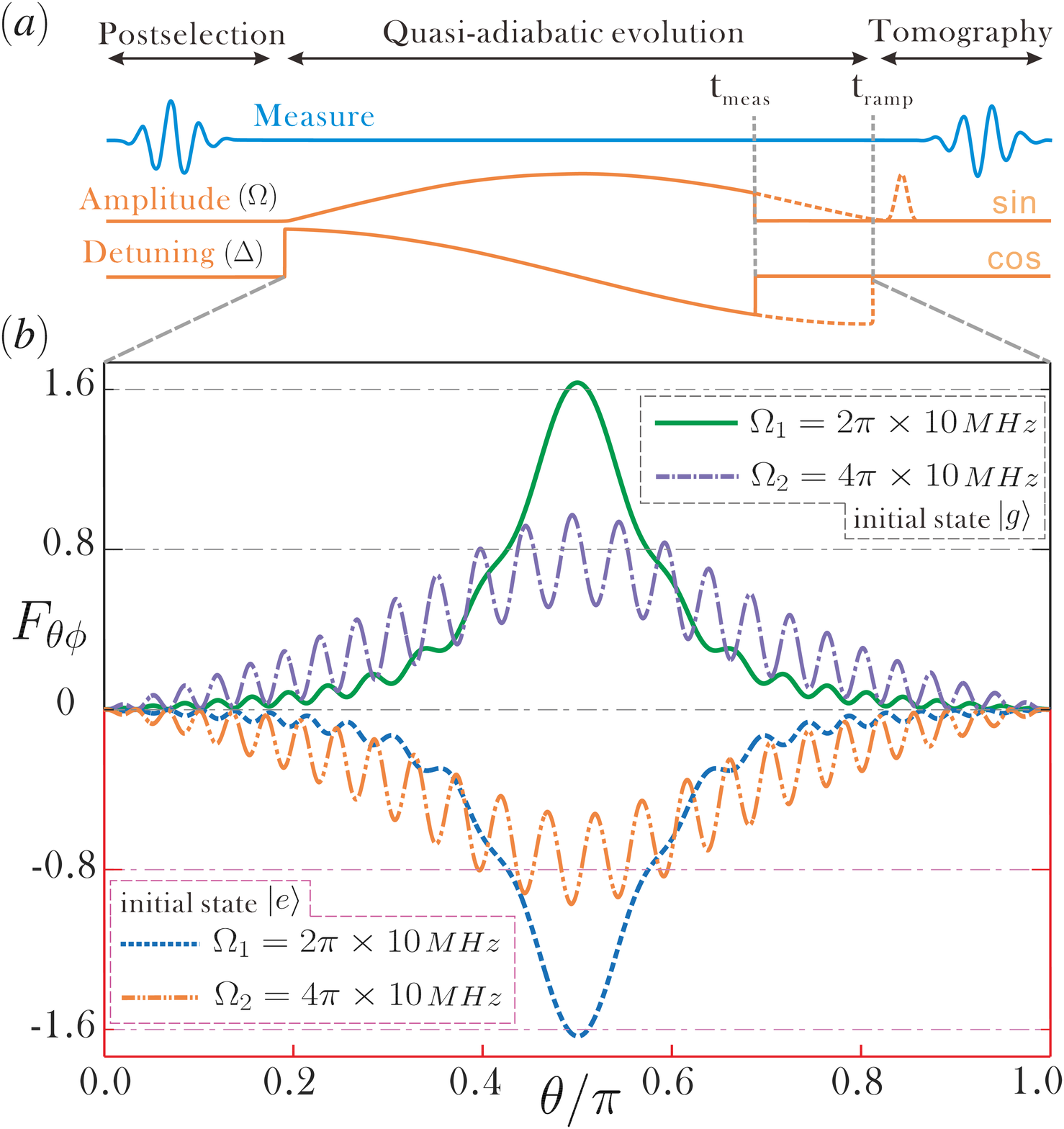}
    \caption{$($Color online$)$ Measuring Berry curvature in a superconducting transmon qubit. (a) Experimental pulse sequence. Following an initial measurement to project into the ground state and the excited state, the detuning and Rabi drive are ramped along Eq.~$($\ref{parameters}$)$, with parameters $\textit{t}_{\textrm{ramp}}$ = 1 $\mu$s, $\Delta_1 = 2\pi\times30$ MHz, $\Omega_1 = 2\pi\times10$ MHz and $\Omega_2 = 4\pi\times10$ MHz. (b) Using Eq.~$($\ref{Measure_Berry_curvature}$)$, one can extract the different Berry curvatures $\textit{F}_{\theta\phi}$ from the measured values of $\langle\textit{g}|\hat{\sigma}_y|\textit{g}\rangle$ (the upper part) and $\langle\textit{e}|\hat{\sigma}_y|\textit{e}\rangle$ (the lower part) at $\Delta_2 = 0$. The curvature of $\Omega_2$ is approximately half that of $\Omega_1$.}
    \label{fig:figure4a_4b}
  \end{figure}

  By changing these parameters $($$\Delta$ and $\Omega$$)$, we can create arbitrary single-qubit Hamiltonians that can be represented in terms of a set of parameters as an ellipsoidal manifold. The eigenstates of this Hamiltonian are
  \begin{eqnarray}\label{eigen_states1}
  |\psi_0\rangle
  &=& \frac{\Omega/2~|0\rangle}{\sqrt{\frac{\Omega^2}{4} + (\textit{E}_0 - \frac{\Delta}{2})^2}} - e^{i\phi}\frac{(\textit{E}_0 - \Delta/2)|1\rangle}{\sqrt{\frac{\Omega^2}{4} + (\textit{E}_0 - \frac{\Delta}{2})^2}},~~~~~~
  \end{eqnarray}
  \begin{eqnarray}\label{eigen_states2}
  |\psi_1\rangle
  &=& \frac{\Omega/2~|0\rangle}{\sqrt{\frac{\Omega^2}{4} + (\textit{E}_1 - \frac{\Delta}{2})^2}} + e^{i\phi}\frac{(\textit{E}_1 - \Delta/2)|1\rangle}{\sqrt{\frac{\Omega^2}{4} + (\textit{E}_1 - \frac{\Delta}{2})^2}},~~~~~~
  \end{eqnarray}
  where $|0\rangle = |\textit{e}\rangle = (1, 0)^{\textit{T}}$ is the excited state and $|1\rangle = |\textit{g}\rangle = (0, 1)^{\textit{T}}$ is the ground state. The corresponding eigenvalues of the eigenstates $|\psi_{1(0)}\rangle$ are ${\textit{E}}_{1(0)} = \pm\frac{1}{2}\sqrt{\Omega^2+\Delta^2}$. We notice that for $\textit{E}_1 = \textit{E}_0$, Eq.~$($\ref{large Berry curvature}$)$ clearly shows that degeneracies are some singular points that will contribute nonzero terms to $\textit{C}_1$ in Eq.~$($\ref{the first Chern number}$)$. In particular, with the choice
  \begin{eqnarray}\label{parameters}
  \Delta = \Delta_1\cos\theta + \Delta_2, ~~~~~~~~~~\Omega = \Omega_n\sin\theta,
  \end{eqnarray}
  the Hamiltonian can be presented in parameter space as an ellipsoidal manifold with cylindrical symmetry about the $z$-axis~\cite{SKKSGVPPL-2014}. Here, we set ellipsoids of size $\Delta_1 = 2\pi\times30$ MHz, and $\Omega_n = 2\textit{n}\pi\times10$ MHz $($\textit{n} = 1, 2, 3$)$, and vary $\Delta_2/(2\pi)$ between $-60$ and $60$ MHz. The topological properties are independent of deformations of the manifold that includes the degenerate point and the choice of these particular parameters does not really matter.
  \begin{figure}[t]
    \centering
    \includegraphics[width=3.2in]{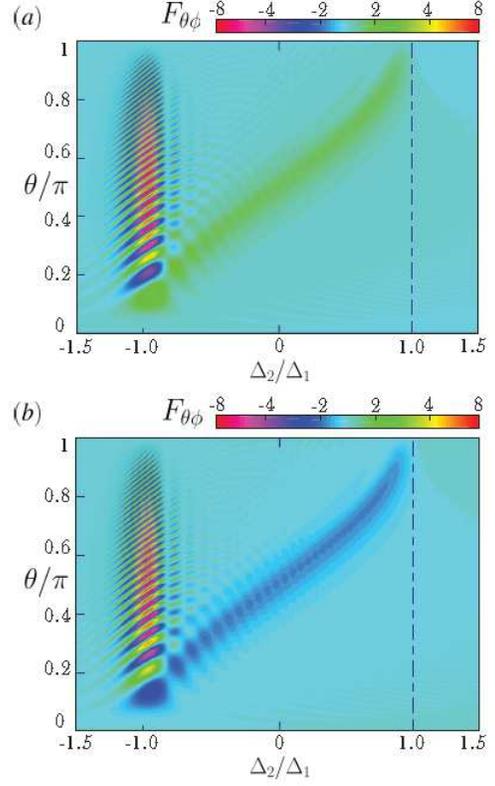}
    \caption{$($Color online$)$ The Berry curvature measured as a function of $\Delta_2/\Delta_1$. In the region of $|\Delta_2/\Delta_1| < 1$, (a) shows the Berry curvature is positive $($the green part$)$, accompanied with the ground state evolution, while (b) shows the curvature is negative $($the blue part$)$, accompanied with the excited state evolution, and it disappears at the dashed line with $\Delta_2/\Delta_1 = 1$ and $\Omega_1 = 2\pi\times10$ MHz.}
    \label{fig:figure5a_5b}
  \end{figure}

  Fig.~\ref{fig:figure4a_4b}$($a$)$ depicts an implementable pulse sequence used to measure the Berry curvature. We respectively initialize the qubit in its bare ground state $|\textit{g}\rangle$ and bare excited state $|\textit{e}\rangle$ at $\theta(\textit{t} = 0) = 0 $ $($this method works for arbitrary eigenstates of the initial Hamiltonian, so the particular state targeted is irrelevant$)$, fix $\phi(\textit{t}) = 0$, and linearly ramp the angle $\theta(\textit{t}) = \pi\textit{t}/\textit{t}_{\textrm{ramp}}$ in time, stopping the ramp at various times $\textit{t}_{\textrm{meas}}\leq\textit{t}_{\textrm{ramp}}$ to execute qubit tomography. From Eq.~$($\ref{large Berry curvature}$)$, the Berry curvature reads
  \begin{eqnarray}\label{Measure_Berry_curvature}
  \textit{F}_{\theta\phi} = \frac{\langle\partial_\phi\hat{H}\rangle}{\upsilon_\theta} = \frac{\Omega_n\sin\theta}{2\upsilon_\theta}\langle\hat{\sigma}_y\rangle,
  \end{eqnarray}
  where $\upsilon_\theta = \dot{\theta}(\textit{t}) = \pi/\textit{t}_{\textrm{ramp}}$.  Fig.~\ref{fig:figure4a_4b}$($b$)$ shows the results of different Berry curvatures with $\Omega_1$ and $\Omega_2$, respectively, for a protocol with $\textit{t}_{\textrm{ramp}} = 1 \mu$s and $\Delta_2 = 0$. We extract the Berry curvatures $\textit{F}_{\theta\phi}$ from the measured values of $\langle\textit{g}|\hat{\sigma}_y|\textit{g}\rangle$ and $\langle\textit{e}|\hat{\sigma}_y|\textit{e}\rangle$.  The Berry curvature is positive when the curving of the surface is elliptic. The sharper the elliptic curving, the greater the Berry curvature. And if the surface starts to show hyperbolic such as a saddle, then the Berry curvature becomes negative, and the sharper the hyperbolic curving of the surface, the smaller the Berry curvature, just the same as the Gauss curvature in Fig.~\ref{fig:figure1}.
  \begin{figure}[t]
    \centering
    \includegraphics[width=3.2in]{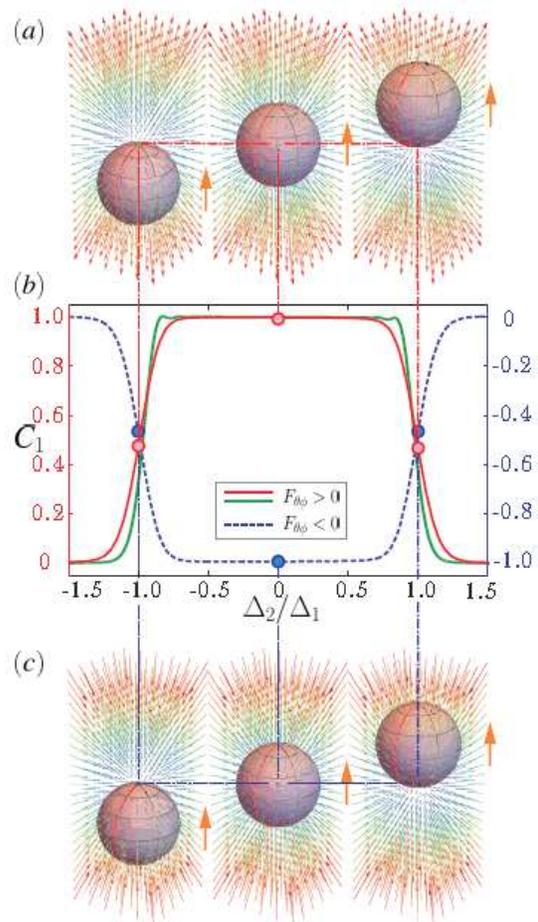}
    \caption{$($Color online$)$ Measuring the topological transition. (a) Fix the degenerate point (green ball) at origin $\Delta = \Omega = 0$, and manipulate the Hamiltonian sphere from down to up, so that the point (act as sources $\textit{g}_{\textit{N}}$) passes from outside to inside and again outside the manifold which releases the magnetic line. (b) The change of $\textit{C}_1$ along with the process of topological transition. The red (real) line corresponds to process (a) and the blue (dashed) line corresponds to process (\textit{c}) with parameters $\Delta_1 = \Omega_3 =  2\pi\times30$ MHz $($the sphere manifold$)$, while the green (real) line corresponds to $\Delta_1 = 2\pi\times30$ MHz and $\Omega_1 = 2\pi\times10$ MHz $($the elliptical manifold$)$, separately ~\cite{SKKSGVPPL-2014}. (\textit{c}) The topological transition is same to that shown in (a), while the point (blue ball) passes from outside to inside and again outside the manifold which acts as sinks $\textit{g}_{\textit{S}}$ that gathers the magnetic line.}
    \label{fig:figure6a_6c}
  \end{figure}

  To induce a topological transition in the qubit, the detuning offset $\Delta_2$ is first changed. At the same time, the ground and excited states evolution are quantitatively modified. But for $|\Delta_2| < |\Delta_1|$, the corresponding Berry curvature as we see in Fig.~\ref{fig:figure5a_5b}$($a$)$ shows the Berry curvature acts like the magnetic field produced by a north magnetic charge $\textit{g}_{\textit{N}}$ $($sources$)$, while Fig.~\ref{fig:figure5a_5b}$($b$)$ shows that it acts like the magnetic field produced by a south magnetic charge $\textit{g}_{\textit{S}}$ $($sinks$)$. The scale of the Berry curvature corresponds to the strength of the magnetic field and it falls with the square of distance between the manifold and the magnetic poles. However, for $|\Delta_2| > |\Delta_1|$, it gives the zero Berry curvature, meaning that the system undergoes a topological transition at $|\Delta_2| = |\Delta_1|$. Such a transition only occurs when the Berry curvature becomes ill defined at the point $\Delta = \Omega = 0$ in Eq.~$($\ref{large Berry curvature}$)$.

  By integrating Eq.~$($\ref{Measure_Berry_curvature}$)$, we obtain the first Chern number
  \begin{eqnarray}\label{Measure_Chern_number}
  \textit{C}_1 = \frac{1}{2\pi}\int_0^{\pi}\textit{d}\theta\int_0^{2\pi}\textit{d}\phi\textit{F}_{\theta\phi} = \int_0^{\pi}\textit{F}_{\theta\phi}\textit{d}\theta.
  \end{eqnarray}
  The measured Chern number $\textit{C}_1$ is plotted in Fig.~\ref{fig:figure6a_6c}$($b$)$, showing a relatively sharp transition at the expected value $|\Delta_2/\Delta_1| = 1$. We find that the topological transition in the elliptical manifold $($the green line$)$ is sharper $($faster$)$ than that in the sphere manifold $($the red line$)$ shown in Fig.~\ref{fig:figure6a_6c}$($b$)$, and it shows that the topological invariant $\textit{C}_1$ is strongly robust against variations in Hamiltonian parameters, such as in Rabi frequency $\Omega_n$ and in detuning $\Delta_1$. The topological transition corresponds to degeneracies moving from outside to inside and again outside the elliptical manifold. In other words, the Chern number is nonzero as long as there exists Berry curvature. From this point of view, we can set up the corresponding relation between topological invariants and magnetic monopoles~\cite{PM-2009,SKKSGVPPL-2014,PR-2014}. Then we can draw such a conclusion, as shown in Fig.~\ref{fig:figure6a_6c}, with a formula~\cite{DCAJ-2004}
  \begin{eqnarray}\label{Final_Chern_number1}
  \textit{C}_1 = \textmd{magnetic number}= \pm1,
  \end{eqnarray}
  where ``1'' is the number of the degeneracy points in parameter space of the Hamiltonian, and the sign ``$\pm$'' corresponds to the polarity of the magnetic charge in parameter space $($$C_1$ = $+1$ $\leftrightarrow$ $\textit{g}_{\textit{N}}$, $C_1$ = $-1$ $\leftrightarrow$ $\textit{g}_{\textit{S}}$$)$.

\section{Results and Discussion}
\label{sec: Results}
   In Eq.~$($\ref{parameters}$)$, $\theta = 0$ and $\pi$ corresponds to $\Delta = \Delta_1+\Delta_2$ and $\Delta = -\Delta_1+\Delta_2$, respectively. For the case with $\Delta = 0$ and $\Omega\neq0$, i.e., the microwave drive induces the resonant transition between the two states $|0\rangle$ and $|1\rangle$ of the qubit, the two eigenstates in Eq.~$($\ref{eigen_states1}$)$ and Eq.~$($\ref{eigen_states2}$)$ become a degenerate state  $|\psi_s\rangle = \frac{1}{\sqrt{2}}(|\textit{e}\rangle + |\textit{g}\rangle)$.

\begin{figure}[t]
    \centering
    \includegraphics[width=3.2in]{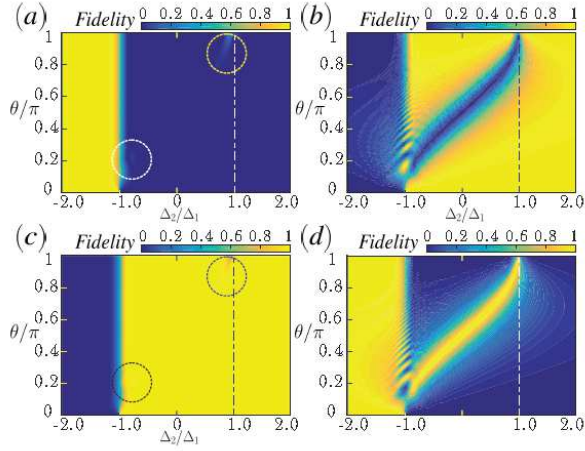}
    \caption{$($Color online$)$ The fidelity of the target states versus $\theta/\pi$ and $\Delta_2/\Delta_1$. The initial state $|\textit{e}\rangle$ $($will evolve within $|\psi_0\rangle$$)$ is set in (a) and (c). The initial state $|\textit{g}\rangle$ $($will evolve within $|\psi_1\rangle$$)$ is set in (b) and (d). In (a) and (b), the density matrix of the target state is $\hat{\rho}(t_f) = |\textit{g}\rangle\langle\textit{g}|$, while in (c) and (d) the density matrix is $\hat{\rho}(t_f) = |\textit{e}\rangle\langle\textit{e}|$. The parameter chosen here are $\Delta_1 = 2\pi\times30$ MHz, $\Omega_1 = 2\pi\times10$ MHz, and $\Delta_2$ ramps from $-2\Delta_1$ to $2\Delta_1$. The Berry curvature only has relatively strong influence around $|\Delta_2/\Delta_1| = 1$ which is shown circled in (a) and (c).}
    \label{fig:figure7a_7d}
  \end{figure}

  Based on this point, we track and investigate the change of the quantum states accompanied with the change of the Berry curvatures. In Fig.~\ref{fig:figure7a_7d}, the fidelity of the target state $|\textit{g}\rangle$ and $|\textit{e}\rangle$ is plotted versus $\theta/\pi$ and $\Delta_2/\Delta_1$, where the fidelity is defined as $\textit{f} = \langle\psi_j|\hat{\rho}(t_f)|\psi_j\rangle$ ($j=0,1$). We note that the quantum state flips at $\Delta_2/\Delta_1 = -1$, when the monopole in parameter space passes from outside to inside the spherical manifold, except the area where the Berry curvatures $($the magnetic fields$)$ exist. However, the quantum state does not flip at $\Delta_2/\Delta_1 \geq 1$, because of the Berry curvatures no longer exist in the manifold and the Gauss theorem of magnetic field turns into Stokes theorem, see in $($a$)$ and $($c$)$ of Fig.~\ref{fig:figure6a_6c}.

  We note that, the strength of the magnetic field $($Berry curvature$)$ has an apparent impact on the quantum state $|\psi_0\rangle$ in $($b$)$ and $($d$)$ of Fig.~\ref{fig:figure7a_7d}, while it only has relatively strong influence on the state $|\psi_1\rangle$ around $|\Delta_2/\Delta_1| = 1$ in $($a$)$ and $($c$)$ of Fig.~\ref{fig:figure7a_7d} $($the dashed circle$)$.

\begin{figure}[t]
    \centering
    \includegraphics[width=3.2 in]{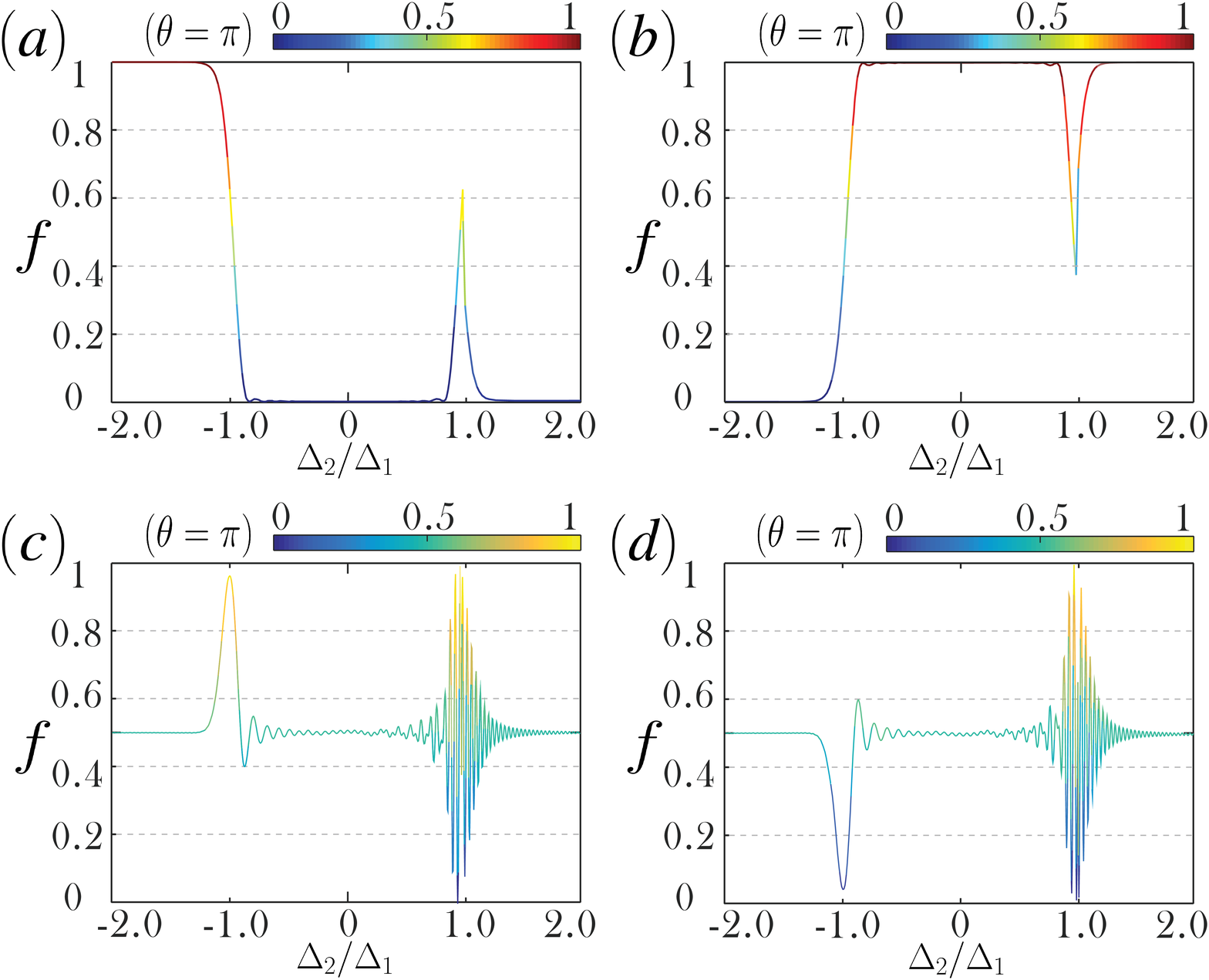}
    \caption{$($Color online$)$ The fidelity of the target states versus $\Delta_2/\Delta_1$ at $\theta = \pi$. (a) The fidelity: $\big|\langle\psi_1|\textit{g}\rangle|^2$ or $\big|\langle\psi_0|\textit{e}\rangle|^2$. (b) The fidelity: $\big|\langle\psi_0|\textit{g}\rangle|^2$ or $\big|\langle\psi_1|\textit{e}\rangle|^2$. (c) The fidelity of the degenerate state: $\big|\langle\psi_0|\psi_s\rangle|^2$. (d) The fidelity of the degenerate state: $\big|\langle\psi_1|\psi_s\rangle|^2$. }
    \label{fig:figure8a_8d}
  \end{figure}

  In order to illustrate the change of the quantum states in the process of topological transition in more detail, we choose a special position at $\theta = \pi$, and thus get $\Delta_1 = \Delta_2$. For such a case, the initial state evolves to the degenerate state $|\psi_s\rangle$. Fig.~\ref{fig:figure8a_8d}$($a$)$ depicts the status of quantum states in $($a$)$ and $($d$)$ of Fig.~\ref{fig:figure7a_7d}, and Fig.~\ref{fig:figure8a_8d}$($b$)$ depicts the status of quantum states in $($b$)$ and $($c$)$ of Fig.~\ref{fig:figure7a_7d}. From Fig.~\ref{fig:figure8a_8d}$($c$)$ and Fig.~\ref{fig:figure8a_8d}$($d$)$, we note that the fidelity is fluctuating around $\Delta_2/\Delta_1 = 1$. We attribute this interesting phenomenon to the influence of the magnetic fields resulting from the magnetic charges. When the charges pass from inside to outside the Hamiltonian manifold, the quantum states influenced by the Berry curvatures will cause ripples in the Hilbert space, a detailed discussion will be presented in the future works. While for the position at $\Delta_2/\Delta_1 = -1$, there is no such apparent fluctuating because the quantum states still have not been affected by the magnetic field. More vividly speaking, the quantum states have not yet been ``magnetized'' by the magnetic monopoles. Actually, according to these phenomena, we find a new way to control the evolution of system quantum states by manipulating $($moving$)$ the monopoles $($degenerate points$)$ in the manifolds.

   Hereinbefore upwards, our main consideration about how to simulate Abelian Wu-Yang monopoles in parameter space just relies on a driven superconducting qubit. However, this general method also could be simulated by other spin-1/2 systems, such as a NMR system in a synthetic magnetic field. A simple experimental scheme is shown in Appendix~{III} of the Supplementary data.

\section{Conclusion}

  We have simulated the Abelian Wu-Yang monopoles in parameter space of the Hamiltonian of a superconducting qubit controlled by a microwave drive for both geometry $($Berry curvature$)$ and topology $($Chern number$)$. The topological structure of the qubit can be captured by the distribution of Berry curvature, which describes the geometry of the eigenstates of the Hamiltonian. We note that during the process of topological transition, the Berry curvature and the fidelity of quantum states have some interesting correlations due to the influence of the magnetic fields resulting from the magnetic charges. We also note that the quantum state flips at the position where the topological transition occurs, when the monopole in parameter space passes from outside to inside and again outside the spherical manifold, except the area where the Berry curvatures $($the magnetic fields$)$ exist. This phenomenon might provide a promising perspective to flexibly manipulate the qubit states by designing the specific synthetic magnetic fields.

  Degenerate points in parameter space of the Hamiltonian act as the sources $($sinks$)$ of $\textit{C}_1$ and are analogues to magnetic monopoles. We also note that the transition of quantum states is asymmetric during the process when the monopole passes from outside to inside and again outside the Hamiltonian manifold. For example, when the monopole passes from inside to outside the Hamiltonian manifold, the quantum states influenced by the Berry curvatures cause ripples in the Hilbert space. However, when the monopole passes from outside to inside the Hamiltonian manifold, there is no such apparent fluctuating. We give a preliminary explanation to this interesting phenomenon by introducing the notion of magnetization of the magnetic charges. This method also can be simulated by other spin-1/2 systems. For example, it can be extended to NMR systems and it is possible to experimentally investigate more intriguing properties of multi-monopoles. This could thus be used to construct new kinds of devices based on synthetic magnetic fields.

\begin{acknowledgments}
  The authors would like to acknowledge insightful discussions with S. B. Zheng. This work was supported by the National Natural Science Foundation of China under Grants  No.11405031, No.11305037, No.11374054, and No.11347114, the Natural Science Foundation of Fujian Province under Grants No.2014J05005, and the fund from Fuzhou University.
\end{acknowledgments}


\begin{thebibliography}{999}

   \bibitem{PAMD-1931} Dirac, P. A. M. Quantised singularities in the electromagnetic field, \textit{Proc. Roy. Soc. A} \textbf{133}, 60 (1931).

   \bibitem{PAMD-1948} Dirac, P. A. M. The theory of magnetic poles, \textit{Phys. Rev.} \textbf{74}, 817 (1948).

   \bibitem{SB-1976} Blaha, S. Quantization rules for point singularities in superfluid $^{3}$He and liquid crystals, \textit{Phys. Rev. Lett.} \textbf{36}, 874 (1976).

   \bibitem{MS-1987} Salomaa, M. M. Monopoles in the rotating superfluid helium-3 A-B interface, \textit{Nature (London)} \textbf{326}, 367 (1987).

   \bibitem{CMS-2008} Castelnovo, C., Moessner, R. \& Sondhi, S. L. Magnetic monopoles in spin ice, \textit{Nature (London)} \textbf{451}, 42 (2008).

   \bibitem{BGCAPF-2009} Bramwell, S. T., Giblin, S. R., Calder, S., Aldus, R., Prabhakaran, D. \& Fennell, T. Measurement of the charge and current of magnetic monopoles in spin ice, \textit{Nature (London)} \textbf{461}, 956 (2009).

   \bibitem{LRPCB-2010} Ladak, S., Read, D. E., Perkins, G. K., Cohen, L. F. \& Branford, W. R. Direct observation of magnetic monopole defects in an artificial spin-ice system, \textit{Nat. Phys.} \textbf{6}, 359 (2010).

   \bibitem{H-1998} Ho, T. L. Spinor Bose condensates in optical traps, \textit{Phys. Rev. Lett.} \textbf{81}, 742 (1998).

   \bibitem{Machida-1998} Ohmi, T. \& Machida, K. Bose-Einstein Condensation with Internal Degrees of Freedom in Alkali Atom Gases, \textit{J. Phys. Soc. Jpn.} \textbf{67}, 1822 (1998).

   \bibitem{PM-2009} Pietil\"{a}, V., \& M\"{o}tt\"{o}nen, M. Creation of Dirac monopoles in spinor Bose-Einstein condensates, \textit{Phys. Rev. Lett.} \textbf{103}, 030401 (2009).

   \bibitem{RPM-2011} Ruokokoski, E., Pietil\"{a}, V. \& M\"{o}tt\"{o}nen, M. Ground-state Dirac monopole, \textit{Phys. Rev. A} \textbf{84}, 063627 (2011).

   \bibitem{RRKMH-2014} Ray, M. W., Ruokokoski, E., Kandel, S., M\"{o}tt\"{o}nen, M. \& Hall, D. S. Observation of Dirac monopoles in a synthetic magnetic field, \textit{Nature (London)} \textbf{505}, 657 (2014).

   \bibitem{RRTMH-2015} Ray, M. W., Ruokokoski, E., Tiurev, K., M\"{o}tt\"{o}nen, M. \& Hall, D. S. Observation of isolated monopoles in a quantum field, \textit{Science} \textbf{348}, 544 (2015).

   \bibitem{TRMHM-2016} Tiurev, K., Ruokokoski, E., M\"{a}kel\"{a}, H., Hall, D. S. \& M\"{o}tt\"{o}nen, M. Decay of an isolated monopole into a Dirac monopole configuration, \textit{Phys. Rev. A} \textbf{93}, 033638 (2016).

   \bibitem{MP-2013} Milde, P. \textit{et al.}, Unwinding of a Skyrmion lattice by magnetic monopoles, \textit{Science} \textbf{340}, 1076 (2013).

   \bibitem{RPF-1982} Feynman, R. P. Simulating physics with computers, \textit{Int. J. Theor. Phys.} \textbf{21}, 467 (1982).


   \bibitem{SSC-1946} Chern, S. S. Characteristic classes of Hermitian manifolds, \textit{Ann. Math.} \textbf{47}, 85 (1946).

   \bibitem{MVB-1984} Berry, M. V. Quantal phase factors accompanying adiabatic changes, \textit{Proc. R. Soc. A} \textbf{392}, 45 (1984).

   \bibitem{DCAJ-2004} Chru\'{s}ci\'{n}ski, D. \& Jamio{\l}kowski, A. \textit{Geometric Phases in Classical and Quantum Mechanics} (Springer Science + Business Media, New York, 2004).

   \bibitem{VGAP-2012} Gritsev, V. \& Polkovnikov, A. Dynamical quantum Hall effect in the parameter space, \textit{Proc. Natl. Acad. Sci. U.S.A.} \textbf{109}, 6457 (2012).

   \bibitem{SKKSGVPPL-2014} Schroer, M. D., Kolodrubetz, M. H., Kindel, W. F., Sandberg, M., Gao, J., Vissers, M. R., Pappas, D. P., Polkovnikov, A. \& Lehnert, K. W. Measuring a topological transition in an artificial spin-1/2 system, \textit{Phys. Rev. Lett.} \textbf{113}, 050402 (2014).

   \bibitem{PR-2014} Roushan, P. \textit{et al.}, Observation of topological transitions in interacting quantum circuits, \textit{Nature (London)} \textbf{515}, 241 (2014).

   \bibitem{YZXYZ-2015} Yang, X. C., Zhang, D. W., Xu, P., Yu, Y. \& Zhu, S. L. Simulating the dynamical quantum Hall effect with superconducting qubits, \textit{Phys. Rev. A} \textbf{91}, 022303 (2015).

   \bibitem{LLLNLPD-2016} Luo, Z. H., Lei, C., Li, J., Nie, X. F., Li, Z. K., Peng, X. H. \& Du, J. F. Experimental observation of topological transitions in interacting multispin systems, \textit{Phys. Rev. A} \textbf{93}, 052116 (2016).

   \bibitem{YCN-1975} Wu, T. T. \& Yang, C. N. Concept of nonintegrable phase factors and global formulation of gauge fields, \textit{Phys. Rev. D} \textbf{12}, 3845 (1975).

   \bibitem{JPGV-1980} Provost, J. \& Vallee, G. Riemannian structure on manifolds of quantum states, \textit{Commun. Math. Phys.} \textbf{76}, 289 (1980).

   \bibitem{MCFL-2010} Ma, Y. Q., Chen, S., Fan, H. \& Liu, W. M. Abelian and non-Abelian quantum geometric tensor, \textit{Phys. Rev. B} \textbf{81}, 245129 (2010).

   \bibitem{MKVGAP-2013} Kolodrubetz, M., Gritsev, V. \& Polkovnikov, A. Classifying and measuring geometry of a quantum ground state manifold, \textit{Phys. Rev. B} \textbf{88}, 064304 (2013).

   \bibitem{PP-2006} Petersen, P. \textit{Riemannian Geometry} (Springer Science + Business Media, New York, 2006).

   \bibitem{MVBJMR-1993} Berry, M. V. \& Robbins, J. M. Chaotic classical and half-classical adiabatic reactions: geometric magnetism and deterministic friction, \textit{Proc. R. Soc.} \textbf{442}, 659 (1993).

   \bibitem{PGDIO-1978} Goddard, P. \& Olive, D. I. Magnetic monopoles in gauge field theories, \textit{Rep. Prog. Phys.} \textbf{41}, 1357 (1978).

   \bibitem{MN-1998} Nakahara, M. \textit{Geometry, Topology and Physics} (Institute of Physics, Bristol, 1998).

   \bibitem{FGCBV-2002} Fuentes-Guridi, I., Carollo, A., Bose, S. \& Vedral, V. Vacuum induced spin-1/2 Berry's phase, \textit{Phys. Rev. Lett.} \textbf{89}, 220404 (2002).

   \bibitem{KYGHSMBDGS-2007} Koch, J., Yu, T. M., Gambetta, J., Houck, A. A., Schuster, D. I., Majer, J., Blais, A., Devoret, M. H., Girvin, S. M. \& Schoelkopf, R. J. Charge-insensitive qubit design derived from the Cooper pair box, \textit{Phys. Rev. A} \textbf{76}, 042319 (2007).

\end{thebibliography}
\end{document}


\textbf{The Title:} Quantum simulation of Abelian Wu-Yang monopoles in spin-1/2 systems

\textbf{The Authors:} Ze-Lin Zhang, Ming-Feng Chen, Huai-Zhi Wu, and Zhen-Biao Yang

\section*{Appendix}
\section{Proof of Eq.~(7) in the main text}

The metric is written in the first fundamental form as
 \begin{eqnarray}\label{the first fundamental form}
 \textit{ds}^2 = \textit{E}(\textit{d}\lambda^\mathbb{1})^2 + \textit{2Fd}\lambda^\mathbb{1} \textit{d}\lambda^\mathbb{2} + \textit{G}(\textit{d}\lambda^\mathbb{2})^2,
 \end{eqnarray}
 then the geometric invariants $($$\textit{K}$, $\kappa_{\textit{g}}$, $\textit{dS}$ and $\textit{dl}$$)$ are given by~\cite{EK-1959,MKVGAP-2013}
 \begin{eqnarray}\label{geometric invariants}
 \begin{split}
 \textit{K} &= \frac{1}{\sqrt{g}}\bigg[\frac{\partial}{\partial\lambda^{\mathbb{2}}}\bigg(\frac{\sqrt{g}~\Gamma^\mathbb{2}_{\mathbb{11}}}{\textit{E}}\bigg) - \frac{\partial}{\partial\lambda^{\mathbb{1}}}\bigg(\frac{\sqrt{g}~\Gamma^\mathbb{2}_{\mathbb{12}}}{\textit{E}}\bigg)\bigg],\\
 \kappa_{\textit{g}} &= \sqrt{g}\textit{G}^{-3/2}\Gamma^{\mathbb{1}}_{\mathbb{22}}, \\
 \textit{dS} &= \sqrt{g}{d}\lambda^{\mathbb{1}} \textit{d}\lambda^{\mathbb{2}},\\
 \textit{dl} &= \sqrt{g}\textit{d}\lambda^{\mathbb{2}},
 \end{split}
 \end{eqnarray}
 where $\kappa_{\textit{g}}$ and $\textit{dl}$ are given for a curve of constant $\lambda^{\mathbb{1}}$. The metric determinant $g = \textit{EG} - \textit{F}^2$,
 and the Christoffel symbols $\Gamma^{\sigma}_{\mu\nu}$ are
 \begin{eqnarray}\label{Christoffel}
 \Gamma^{\sigma}_{\mu\nu} = \frac{1}{2}{\textit{g}}^{\sigma\lambda}(\partial_{\nu}~{\textit{g}}_{\mu\lambda} + \partial_{\mu}~{\textit{g}}_{\nu\lambda} - \partial_{\lambda}~{\textit{g}}_{\mu\nu}),
 \end{eqnarray}
 where $\textit{g}^{\mu\nu}$ is the inverse of the metric tensor $\textit{g}_{\mu\nu}$.

 Now we consider a two-level quantum system living in a complex space $\mathds{C}^2$. The corresponding space of quantum state may be parameterized by standard spherical angles $($$\theta,\phi$$)$. Because of the ellipsoid and sphere are homeomorphism to each other. For the sake of simplicity, we take a local section: $\textit{S}^2\ni \theta, \phi \rightarrow |\psi(\theta, \phi)\rangle\in \textit{S}^3 = \textit{S}(\mathds{C}^2)$, defined by
 \begin{eqnarray}\label{qubit}
 |\psi_{1}(\theta, \phi)\rangle = \cos({\theta}/{2})|0\rangle + \textit{e}^{i\phi}\sin({\theta}/{2})|1\rangle,
 \end{eqnarray}
 where we set $\cos({\theta}/{2}) = \frac{\Omega}{2}\big/{\sqrt{\frac{\Omega^2}{4} + (\textit{E}_1 - \frac{\Delta}{2})^2}}$ and $\sin({\theta}/{2}) = (\textit{E}_1 - \frac{\Delta}{2})\big/{\sqrt{\frac{\Omega^2}{4} + (\textit{E}_1 - \frac{\Delta}{2})^2}}$.

 Let $\lambda^{\mathbb{1}} = \theta$ and $\lambda^{\mathbb{2}} = \phi$, one can easily compute the following:
 \begin{eqnarray}\label{metric components}
 \begin{split}
 &\langle\partial_{\theta}\psi_{1}|\partial_{\theta}\psi_{1}\rangle = 1/4,~~~\langle\partial_{\phi}\psi_{1}|\partial_{\phi}\psi_{1}\rangle = \sin^2({\theta}/{2}),\\
 &\langle\partial_{\theta}\psi_{1}|\partial_{\phi}\psi_{1}\rangle = -\langle\partial_{\phi}\psi_{1}|\partial_{\theta}\psi_{1}\rangle\ = i/4\sin\theta, \\
 &\langle\partial_{\phi}\psi_{1}|\psi_{1}\rangle\langle\psi_{1}|\partial_{\phi}\psi_{1}\rangle = \sin^4({\theta}/{2}),\\
 &\langle\psi_{1}|\partial_{\phi}\psi_{1}\rangle = -\langle\partial_{\phi}\psi_{1}|\psi_{1}\rangle\ = i\sin^2({\theta}/{2}).\\
 \end{split}
 \end{eqnarray}
 Hence, the corresponding components of the Fubini-Study metric tensor in Eq.~$($3$)$ in the main text are given by
 \begin{eqnarray}\label{Fubini-Study}
 \begin{split}
 \textit{E} &= \textit{g}_{\theta\theta} = 1/4,~~~~\textit{F} = \textit{g}_{\theta\phi} = 0,\cr
 \textit{G} &= \textit{g}_{\phi\phi} = 1/4 \sin^2({\theta}/{2}),\cr
 \end{split}
 \end{eqnarray}
 with $\textit{ds}^2 = 1/4\textit{d}\theta^2 + 1/4\sin^2\theta\textit{d}\phi^2$. For example, the metric agrees with the standard metric tensor on a sphere of radius $r =$ 1/2.
 We rewrite the Fubini-Study metric tensor in matrix form
 \begin{eqnarray}\label{g ij}
 \textit{g}_{\mu\nu} = \left( \begin{array}{cc}
                    1/4 & 0 \\
                    0 & 1/4 \sin^2{\theta}  \\
                  \end{array}
 \right),~
 \textit{g}^{\mu\nu} = \left( \begin{array}{cc}
                    {4} & 0 \\
                    0 & {4}\csc^2{\theta}  \\
                  \end{array}
 \right),
 \end{eqnarray}
 and get
 \begin{eqnarray}\label{partial g ij}
 \partial_\theta\textit{g}_{\mu\nu} = \left( \begin{array}{cc}
                    0 & 0 \\
                    0 & 1/2\sin{\theta}\cos\theta  \\
                  \end{array}
 \right),
 \partial_\phi\textit{g}^{\mu\nu} = \left( \begin{array}{cc}
                    0 & 0 \\
                    0 & 0 \\
                  \end{array}
 \right),
 \end{eqnarray}
 where $\mu$ $($$\nu$$)$ = $\theta$, $\phi$. Now we substitute Eq.~$($\ref{partial g ij}$)$ into Eq.~$($\ref{Christoffel}$)$, which shows that
 \begin{eqnarray}\label{Christoffel2}
 \begin{split}
 \Gamma^{\theta}_{\phi\phi}
 &= 1/2{\textit{g}}^{\theta \theta}(2\partial_{\phi}~{\textit{g}}_{\phi \theta} - \partial_{\theta}~{\textit{g}}_{\phi\phi})= -\cos\theta\sin\theta, \\
 \Gamma^{\phi}_{\phi\theta}&= 1/2{\textit{g}}^{\phi \phi}(\partial_{\theta}~{\textit{g}}_{\phi \phi})= \cot\theta.
 \end{split}
 \end{eqnarray}
 The corresponding matrix form is
 \begin{eqnarray}\label{Christoffel matrix}
 \Gamma^{\theta}_{\mu\nu} = \left( \begin{array}{cc}
                    0 & 0 \\
                    0 & -\cos{\theta}\sin{\theta}  \\
                  \end{array}
 \right),
 \Gamma^{\phi}_{\mu\nu} = \left( \begin{array}{cc}
                    0 & \cot\theta \\
                    \cot\theta & 0 \\
                  \end{array}
 \right).
 \end{eqnarray}
 Then we calculate the geometric invariants appearing in Eq.~$($\ref{geometric invariants}$)$
 \begin{eqnarray}\label{geometric invariants2}
 \begin{split}
 \sqrt{g}
 &= \sqrt{\textit{EG}-\textit{F}^2} = 1/4\sin\theta,~~~~\textit{K}= -\frac{1}{g}\frac{\partial}{\partial\theta}\bigg(\frac{\sqrt{g}~\Gamma^{\phi}_{\theta\phi}}{\textit{E}}\bigg) = 4,\\
 \kappa_{\textit{g}}
 &= \sqrt{g}\textit{G}^{-3/2}\Gamma^{\theta}_{\phi\phi}=-2\cot\theta,~~~~\textit{dS}= 1/4\sin\theta\textit{d}\theta\textit{d}\phi,~~~~\textit{dl}= 1/2\sin\theta\textit{d}\phi.
 \end{split}
 \end{eqnarray}
 where we set $r = 1/2$. The integer Euler characteristic for a sphere is $\chi(\textit{S}^2) = 2$, and Eq.~$($6$)$ in the main text satisfies
 \begin{eqnarray}\label{Gauss-Bonnet2}
 \iint_{{\mathcal{S}}^2}\textit{K}~\textit{dS} + \oint_{\partial{{\mathcal{S}}^2}}\kappa_{\textit{g}}~\textit{dl} &=& 4\pi,
 \end{eqnarray}
 where we note that $\theta=\pi/2$ and $\kappa_{\textit{g}}=0$, thus to finally we get
 \begin{eqnarray}\label{Global Gauss-Bonnet2}
 \frac{1}{2\pi}\oint_{{\mathcal{S}}^2}\textit{K}~\textit{dS} = \chi({\mathcal{S}}^2).
 \end{eqnarray}

\section{The first Chern number and Wu-Yang monopole}
\label{sec: The first Chern number}

  Let us first introduce the Chern classes and consider a complex hermitian vector bundle with the structure Lie group $\textit{G}=\textit{GL}(\textit{n},\mathds{C})$. We define the invariant polynomials $\textit{c}_k\in \textit{I}_k(\textit{G})$ by
  \begin{equation}\label{invariant_polynomials}
  \textmd{det}\bigg(\mathds{1}+\frac{i}{2\pi}\textit{F}\bigg)=: \overset{\textit{n}}{\underset{k=0}{\sum}}\textit{c}_k(\textit{F}).
  \end{equation}
  One calls $\textit{c}_k(\textit{F})$ a $\textit{k}$-th Chern form. The $\textit{k}$-th Chern class $\textit{C}_k(\mathcal{P})$ of the bundle $\mathcal{P}$ is defined by
  \begin{equation}\label{k_Chern_class1}
  \textit{C}_k(\mathcal{P}):= \llbracket\textit{c}_k(\textit{F})\rrbracket,
  \end{equation}
  where ``$\llbracket$ $\rrbracket$'' represents the cohomology class. It is easy to show that
  \begin{equation}\label{k_Chern_class2}
  \textit{c}_k(\textit{F})= \frac{(-1)^k}{(2\pi i)^k}~\epsilon_{i_1\cdots i_k}^{j_1\cdots j_k}~\textit{F}_{~j_1}^{i_1}\wedge\cdots\wedge \textit{F}_{~j_k}^{i_k},
  \end{equation}
  and
  \begin{equation}\label{square_bracket}
  \epsilon_{i_1\cdots i_k}^{j_1\cdots j_k} := \delta_{~[i_1} ^{j_1}\cdots\delta_{~i_k]}^{j_k},
  \end{equation}
  where ``$\wedge$'' stands for the wedge product and square bracket for anti-symmetrization, for example, $\alpha_{[i} ^{j}\beta_{i]}^{j}=\alpha_{[i}\beta_{j]}=(\alpha\wedge\beta)_{ij}=\alpha_i\beta_j-\alpha_j\beta_i$.
  One easily finds that
  \begin{eqnarray}\label{bundle_class}
  \textit{C}_0(\mathcal{P})&=&1,\cr
  \textit{C}_1(\mathcal{P})&=&\frac{i}{2\pi}\textmd{Tr}~\textit{F},\cr
  \textit{C}_2(\mathcal{P})&=&\frac{1}{2}\bigg(\frac{i}{2\pi}\bigg)^2\big(\textmd{Tr}~\textit{F}\wedge\textmd{Tr}~\textit{F}-\textmd{Tr}(\textit{F}\wedge\textit{F})\big),\cr
  &\vdots&\cr
  \textit{C}_{\textit{n}}(\mathcal{P})&=&\bigg(\frac{i}{2\pi}\bigg)^{\textit{n}} \textmd{det}~\textit{F}.
  \end{eqnarray}
  For example, $\textit{U}(1)$-bundles are characterized by the first Chern class $\textit{C}_1(\mathcal{P})$:
  \begin{equation}\label{square_bracket}
  \textit{C}_1(\mathcal{P})=\frac{i}{2\pi}\textmd{Tr}~\textit{F}=\frac{i}{2\pi}\textit{F}.
  \end{equation}

  Let $\mathcal{P}\rightarrow \mathcal{M}$ be a principal $\textit{G}$-bundle and suppose that the base $\mathcal{M}$ is an oriented compact manifold of dimensional 2$\textit{n}$. Then the value of the integral
  \begin{equation}\label{bundle_Chern_number}
  \int_{\mathcal{M}} \textit{C}_{\textit{n}}(\mathcal{P})=\int_{\mathcal{M}} \textit{c}_{\textit{n}}(\textit{F}),
  \end{equation}
  is called the Chern number of the bundle, to speak of, this abstract mathematical concept plays an important role in modern physics. To deduce the quantization of $\textit{C}_1$, we will use an argument similar to Dirac's argument showing that the magnetic monopole charge is quantized~\cite{PAMD-1931,PAMD-1948}. After that, we will explicitly relate this quantized value to the number of enclosed ``magnetic monopoles'' in parameter space~\cite{DCAJ-2004}. Following Wu and Yang~\cite{YCN-1975}, we let $\textit{U}_{\textit{N}}$ and $\textit{U}_{\textit{S}}$ be open subsets in $\mathcal{S}^2$ such that:

  $1.$ $\textit{U}_{\textit{N}}$ $($$\textit{U}_{\textit{S}}$$)$ contains the north $($south$)$ pole;

  $2.$ $\textit{U}_{\textit{N}}\cup \textit{U}_{\textit{S}}=\mathcal{S}^2$, and $\textit{U}_{\textit{N}}\cup \textit{U}_{\textit{S}}\neq {\O}$.\\
  Now, a $\textit{U}(1)$-bundle over $\mathcal{S}^2$ is uniquely determined by a transition function
  \begin{equation}\label{transition_function}
  \Gamma_{\textit{NS}}:\textit{U}_{\textit{N}}\cup \textit{U}_{\textit{S}}\rightarrow \textit{U}(1).
  \end{equation}
  For $\textit{G}=\textit{U}(1)$, we have electromagnetism. A local transformation $\textit{T}$ induces the following transformations of local connections and curvatures:
  \begin{eqnarray}\label{locla_gauge_transformation1}
  \textit{A}'= {\textit{T}}^{-1}\cdot \textit{A}\cdot \textit{T}+{\textit{T}}^{-1}\cdot \textit{dT},~~~~~~
  \textit{F}'= {\textit{T}}^{-1}\cdot\textit{F}\cdot\textit{T}.
  \end{eqnarray}
  Let us take $\textit{T}=\Gamma_{\textit{NS}}$($\theta,\phi$)$:={\textit{e}}^{i\textit{n}\phi}$. Clearly, the local connection forms are related by
  \begin{equation}\label{connection_form}
  \textit{A}_{\textit{S}}=\textit{A}_{\textit{N}}+\Gamma_{\textit{NS}}^{-1}\textit{d}~\Gamma_{\textit{NS}}=\textit{A}_{\textit{N}}+\textit{ind}\phi,
  \end{equation}
  and $\textit{F}_{\textit{N}}=\textit{F}_{\textit{S}}$.
  Let us compute the corresponding Chern number of the magnetic bundle $($$\textit{U}(1)$-bundle$)$:
  \begin{equation}\label{Chern_magnetic}
  \textit{C}_1=\iint_{\mathcal{S}^2}{\textit{C}}_1(\mathcal{P}).
  \end{equation}

  \begin{figure}[t]
    \centering
    \includegraphics[width=2.3in]{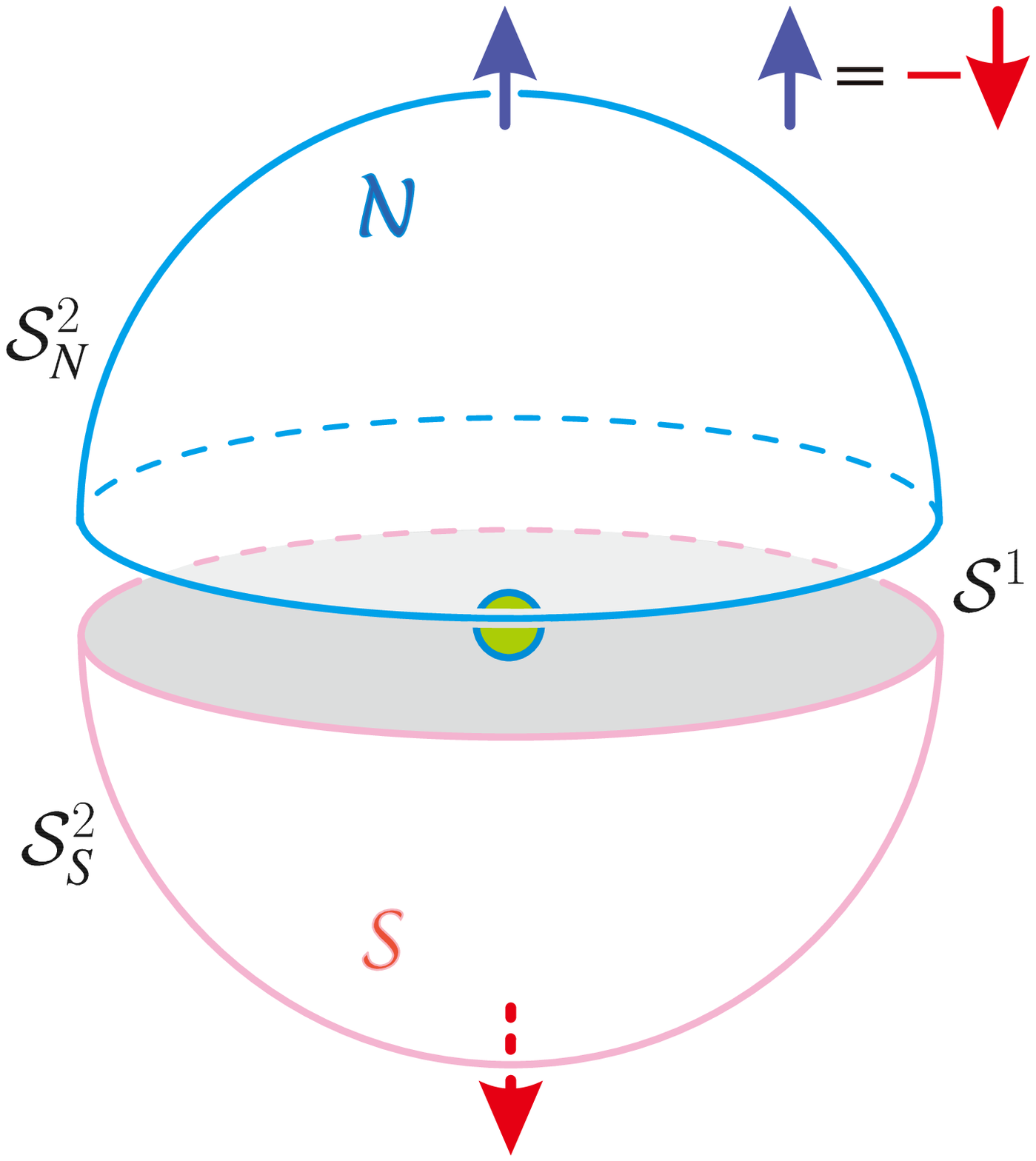}
    \caption{$($Color online$)$ Wu-Yang monopole $($the green ball$)$. We may avoid singularities of Dirac string by adopting $\vec{\textit{A}}_{\textit{N}}$ in the northern hemisphere $\mathcal{S}^2_{\textit{N}}$ and $\vec{\textit{A}}_{\textit{S}}$ in the southern hemisphere $\mathcal{S}^2_{\textit{S}}$ of the sphere ${\mathcal{S}^2}$ surrounding the monopole. $\mathcal{S}^1$ is the oriented boundary of surface.}\label{monopole}
    \label{fig:figure9}
  \end{figure}

  To do this, let us define up, $\mathcal{S}_{\textit{N}}^2$, and down, $\mathcal{S}_{\textit{S}}^2$, hemispheres, as is shown in Fig.~\ref{fig:figure9}, as follows:
  \begin{eqnarray}\label{hemisphere}
  \begin{split}
  \mathcal{S}_{\textit{N}}^2
  &:=&\big\{({\lambda}^{\mathbb{1}},{\lambda}^{\mathbb{2}},{\lambda}^{\mathbb{3}})\in{\mathcal{S}^2}~\big|~{\lambda}^{\mathbb{3}}\geq0\big\},\\
  \mathcal{S}_{\textit{S}}^2
  &:=&\big\{({\lambda}^{\mathbb{1}},{\lambda}^{\mathbb{2}},{\lambda}^{\mathbb{3}})\in{\mathcal{S}^2}~\big|~{\lambda}^{\mathbb{3}}\leq0\big\}.
  \end{split}
  \end{eqnarray}
  Evidently, $\mathcal{S}_{\textit{N}}^2\cup\mathcal{S}_{\textit{S}}^2=\mathcal{S}^2$ and $\mathcal{S}_{\textit{N}}^2\cap\mathcal{S}_{\textit{S}}^2=\mathcal{S}^1$, where $\mathcal{S}^1$ is an ``equatorial'' circle that we provide with the orientation it inherits from $\mathcal{S}_{\textit{N}}^2$. One has
  \begin{eqnarray}\label{on}
  \begin{split}
  \textit{F}_{\textit{N}}= d\textit{A}_{\textit{N}}~~\textmd{on}~~\mathcal{S}_{\textit{N}}^2\subset \textit{U}_{\textit{N}},~~~~
  \textit{F}_{\textit{S}}= d\textit{A}_S~~~\textmd{on}~~\mathcal{{\textit{S}}}_{\textit{S}}^2\subset \textit{U}_{\textit{S}},
  \end{split}
  \end{eqnarray}
  and, since $\textmd{Tr}~\textit{F}=\textit{F}$, the Stokes theorem (pay attention to the orientations) implies that
  \begin{eqnarray}\label{Stokes_monopole1}
  \begin{split}
  \iint_{\mathcal{S}_{\textit{N}}^2}\textit{F}_{\textit{N}}&=\iint_{\mathcal{S}_{\textit{N}}^2}d\textit{A}_{\textit{N}}=\oint_{\mathcal{S}^1}\textit{A}_{\textit{N}},~~~~
  \iint_{\mathcal{S}_{\textit{S}}^2}\textit{F}_{\textit{S}}&=\iint_{\mathcal{S}_{\textit{S}}^2}d\textit{A}_{\textit{S}}=-\oint_{\mathcal{S}^1}\textit{A}_{\textit{S}}.
  \end{split}
  \end{eqnarray}
  Thus we obtain the following formula for the Chern number:
  \begin{eqnarray}\label{final_Chern}
  \iint_{\mathcal{S}^2}C_1(\mathcal{P})
  &=&
  \frac{i}{2\pi}\iint_{\mathcal{S}_{\textit{N}}^2}\textit{F}_{\textit{N}}+\frac{i}{2\pi}\iint_{\mathcal{S}_{\textit{S}}^2}\textit{F}_{\textit{S}}\cr \cr
  &=&
  \frac{i}{2\pi}\oint_{\mathcal{S}^1}(\textit{A}_{\textit{N}}-\textit{A}_{\textit{S}})
  =\frac{i}{2\pi}\oint_{\mathcal{S}^1}(-\textit{ind}\phi)=\textit{n}.
  \end{eqnarray}
  Note that the Chern number does not depend on a particular choice of $\textit{A}$ but only on the transition function $\Gamma_{\textit{NS}}$, which uniquely defines the monopole bundle. Any two bundles with the same~\cite{DCAJ-2004}
  \begin{eqnarray}\label{Final_Chern_number}
  \textmd{magnetic number}=\textmd{Chern number}=\textit{n}.
  \end{eqnarray}

\section{An experimental scheme of simulating monopoles by synthetic magnetic fields}
\label{sec: synthetic magnetic fields}

  \begin{figure}[h]
    \centering
    \includegraphics[width=2.5in]{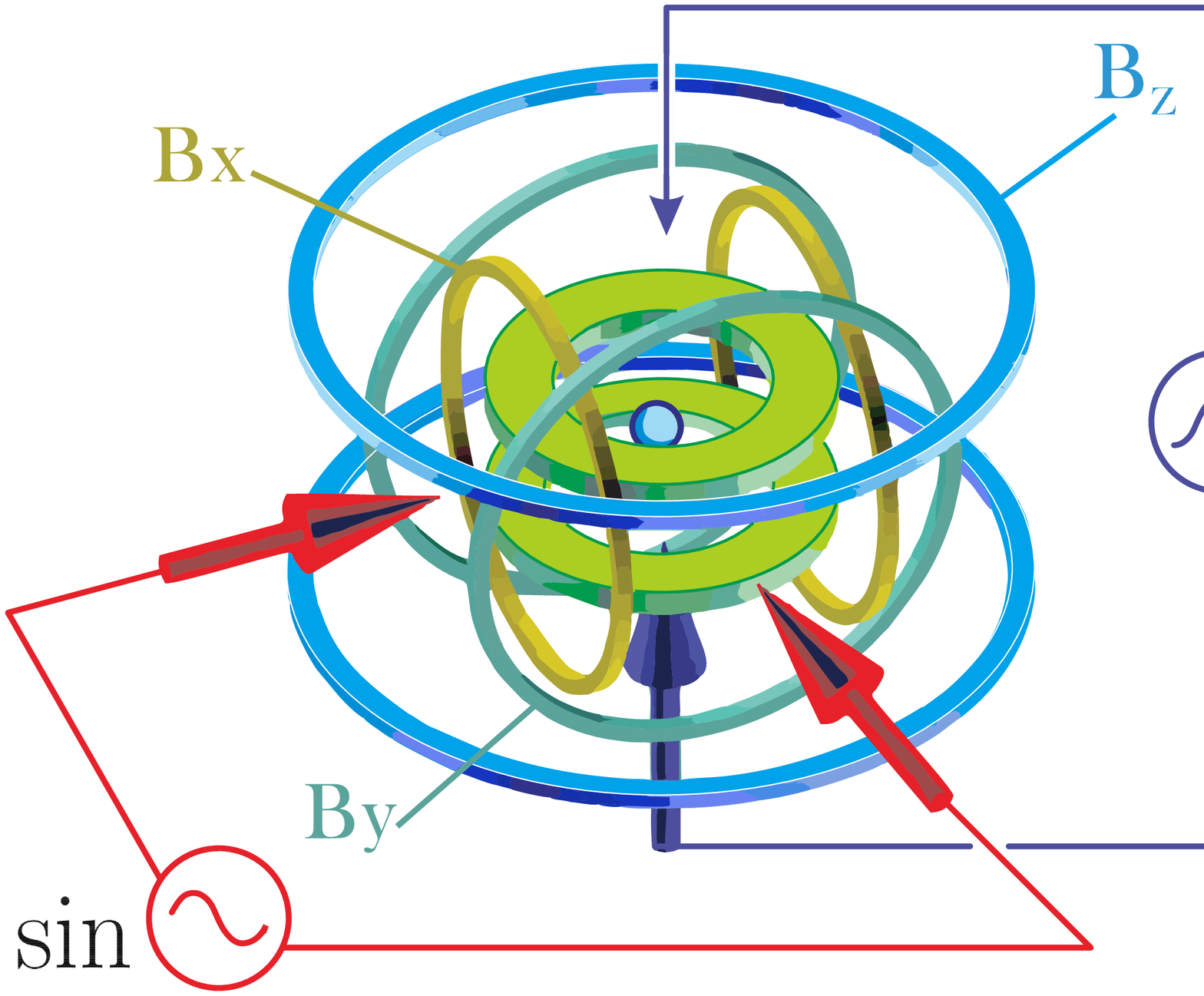}
    \caption{$($Color online$)$ The experiment implementation for synthetic magnetic fields. Experimental set-up shows synthetic magnetic charge $($the light blue ball$)$ and bias field $($$\textit{B}_x$, $\textit{B}_y$ and $\textit{B}_z$$)$ coils. Red arrows show horizontal magnetic fields cyclical change in the form of function of sine, and purple arrow indicates vertical magnetic field cyclical change in the form of function of cosine. $($Modified experimental apparatus appears on Ref.~\cite{RRKMH-2014}$)$}
    \label{fig:figure10}
  \end{figure}

  Now we consider an electron moving in a rotating synthetic magnetic field
  \begin{eqnarray}\label{B field}
  \vec{\textit{B}} = \textit{B}_x \cos\phi\vec{\textit{e}}_x + \textit{B}_y \cos\phi\vec{\textit{e}}_y + \textit{B}_z\vec{\textit{e}}_z.
  \end{eqnarray}
  and the Hamiltonian can be written as $($$\hbar \equiv 1$$)$
  \begin{eqnarray}\label{H of B}
  \hat{H} = \textit{M}_b[\textit{B}_x \cos\phi\hat{\sigma}_x + \textit{B}_y \cos\phi\hat{\sigma}_y + \textit{M}_b\textit{B}_z\hat{\sigma}_z].
  \end{eqnarray}
  where $\textit{M}_b = \hbar e/(2m)$ is the Bohr magneton, and $\textit{B}_x = \textit{B}_y = \textit{B}_1\sin\theta$ and $\textit{B}_z = \textit{B}_1\cos\theta + \textit{B}_2$ are the magnitude of the horizontal magnetic fields and the magnitude of the vertical magnetic field, separately. $\theta$ and $\phi = \omega_0\textit{t}$ are the phase of the drive magnetic field and the frequency of the magnetic field, separately. From this point of view, the magnetic monopoles could be implemented in the real physical systems, as shown in Fig.~\ref{fig:figure10}.